\documentclass[
  aps, physrev,        
  reprint,             
  superscriptaddress,  
  longbibliography,    
  floatfix             
]{revtex4-2}
\usepackage{graphicx} 

\usepackage{dcolumn}
\usepackage{multirow}
\usepackage{bm}

\usepackage{amsmath,amssymb}
\usepackage{comment}

\begin{document}

\title{Dimensional crossover in ultradilute Bose-Bose liquid mixtures}

\author{L. Vranje\v{s} Marki\'c}
\email{leandra@pmfst.hr} 
\affiliation{University of Split, Faculty of Science, Split, Croatia}

\author{I. Popari\'c}
\affiliation{University of Split, Faculty of Science, Split, Croatia}
\affiliation{Departament de F\'isica, Universitat Polit\`ecnica de Catalunya, Barcelona, Spain}
\affiliation{University of Zagreb Faculty of Science, Zagreb, Croatia}

\author{K. D\v{z}elalija}
\affiliation{University of Split, Faculty of Science, Split, Croatia}

\author{P. Stipanovi\'c}
\affiliation{University of Split, Faculty of Science, Split, Croatia}

\author{J. Boronat}
\affiliation{Departament de F\'isica, Universitat Polit\`ecnica de Catalunya, Barcelona, Spain}

\date{November 5, 2026}

\begin{abstract}

We study ultradilute $^{39}$K Bose-Bose bulk mixtures and 
droplets in an external harmonic potential that confines them in one spatial 
direction towards the two-dimensional (2D) limit.  Equations of state for 
several confinements are obtained with quantum Monte Carlo (QMC)
at $T=0$, using interaction potentials that include information on the $s$-wave 
scattering length $a$ and the effective range $r_{\rm eff}$. Performing the 
calculations using two different interaction potential models we have determined 
the range of confinements for which equations of state are universal in terms of 
$a$ and $r_{\rm eff}$. Based on the QMC equation of 
state, we develop a 2D QMC density functional for each confinement strength and 
use it together with the local density approximation to determine properties of 
the self-bound drops.   For moderate squeezing, energies and droplet 
profiles obtained using the 2D QMC functional agree well with those obtained 
using 3D functionals, while offering a substantial reduction in computational 
cost, and a consistent approach in crossover to 2D. Noticeably, our results 
approach 2D mean-field (MF) + Lee-Huang-Yang (LHY) predictions only for the 
most strongly confined systems for which universality in terms of $a$ and 
$r_{\rm eff}$ is observed. This implies a very narrow range of confinements 
for 
which 2D LHY functionals are applicable, which affects 
the study of vortices using 2D functionals.

\end{abstract}

\maketitle
\section{\label{sec:I}Introduction}
Ultradilute self-bound Bose-Bose droplets, predicted by 
Petrov~\cite{petrov2015quantum}, are stable due to beyond mean-field (MF) effects, 
the most important one being the Lee-Huang-Yang correction 
(LHY)~\cite{lee1957eigenvalues,LARSEN196389}. So far, they have been realized 
in several mixtures with attractive interspecies and repulsive intraspecies 
interactions, beginning with mixtures formed by two hyperfine states of 
$^{39}$K~\cite{Cabrera,semeghini2018self} and afterwards in heterogeneous 
Bose-Bose droplets of $^{41}$K-$^{87}$Rb and 
$^{23}$Na-$^{87}$Rb~\cite{derrico2019observation,Guo21,Cavicchioli}. 
Ultradilute droplets had been observed earlier in confined dipolar 
atoms~\cite{kadau,ferrier,schmitt,chomaz}, arising from the asymetry between 
repulsion and attraction of dipoles depending on their position.
 Recently, 
the formation of self-bound clusters was also theoretically predicted~\cite{prl_molecules} and observed~\cite{Will} 
 in ultracold dipolar molecules.

Dimensionality significantly affects the droplet stability. While it takes a 
critical number of atoms to form a self-bound droplet in three dimensions (3D), 
self-bound droplets in two dimensions (2D) are formed whenever the intraspecies 
interactions are repulsive and the interspecies ones are attractive~\cite{AP}. 
The equation of state including the 2D LHY contribution~\cite{AP}, which is 
expressed as a function of the 2D scattering lengths $a_{\rm 2D}$, is then used 
to study quasi-2D systems, via the connection  $a_{\rm 2D}=2e^{-\gamma}\sqrt{\pi 
a_z^2/A}\exp[-\sqrt{\pi /2}a_z/a_{\rm 3D}]$, where $a_{\rm 3D}$ is the 3D 
scattering length, $a_z$ the harmonic oscillator length in the direction of 
confinement and $A\approx 0.905$~\cite{Petrov2001}. This approach is widely 
used 
in the study of vortices and excitations in 
quasi-2D droplets~\cite{Li2018,tengstrand_rotating_2019,Kartashov19,
kartashov_2020, 
examilioti_ground_2020,Dong21,Sturmer21,nikolaou_rotating_2023-1,gu_self-bound_2023, 
nikolaou_rotating_2024,Fei2024,Jha2025,Paredes2025,Paredes2025-1,Wang2025}. However, 
the extension of the LHY correction from 3D to 2D is non-trivial. 
Beyond MF effects were in fact studied in the crossover from three to 
two dimensions~\cite{Ilg,Zin18}, resulting in integral expressions for the LHY 
term and conditions for the use of 
3D local-density functionals to study droplets in 
confinement. Droplets in a quasi-2D setup created by an external harmonic
potential in one spatial direction have also been studied using the modified 
gapless Hartree-Fock-Bogoliubov
method~\cite{Zin_2022, Zin_2022_PRA}, while recent work has been directed to 
droplets in tight transversal box potential of length $L_z$,~\cite{Pelayo}, 
using the extended Gross-Pitaevskii equation with the LHY term from 
Ref.~\cite{Zin18}.  In this latter study, it is found  
that the quasi-2D approach starts to depart from results using the 2D LHY 
expression already at very tight traps. 

The study of 3D droplets has shown effects beyond the MF+LHY 
theory in the equation of state of the 
liquid,~\cite{Staudinger,cikojevic2019PRA}, the critical minimum
number of atoms to achieve self binding~\cite{cikojevic2020NJP},  and the  
excitation 
modes~\cite{cikojevic}. They appear more pronounced in quasi-2D, as observed in 
the confined $^{39}$K droplets~\cite{Cabrera}, where the observed critical 
numbers and sizes of droplets did not match the prediction of MF+LHY theory. 
Much better agreement was achieved by using a functional based on quantum Monte 
Carlo (QMC) calculations of the bulk phase, with interactions described by both 
the scattering length and the effective range~\cite{cikojevic2020NJP}, showing 
that two scattering parameters are needed for a universal description.

In the present work,  we perform diffusion Monte Carlo (DMC)
calculations of bulk squeezed mixture in the 3D to 2D crossover regime and, 
from 
the obtained equations of state, build 2D QMC density functionals, which are 
then 
used to study the droplets. Our interaction potential models include the effects 
of the effective range, allowing us to explore the extent of the universality in 
terms of two scattering parameters as the liquid is squeezed towards 2D. Our 
results clearly show that the use of the 2D LHY functional~\cite{AP} in 
quasi-2D systems 
has a limited applicability to very tight confinement. The present 2D QMC functional is instead able to perform much better and in a more extended range 
of squeezing.

The paper is organized as follows. In Sec. \ref{sec:II} we define the system under consideration and methods used to study it. In Sec. \ref{sec:III} we present and discuss our findings, while Sec. \ref{sec:IV} gives the summary of results and conclusion.

\section{\label{sec:II}Method}
We consider a mixture of $^{39}$K isotopes in the external 
harmonic potential $V_\mathrm{ext}$ in the $z$ direction, described by the 
3D Hamiltonian
    \begin{equation}
        H = \frac{-\hbar^2}{2m}\sum_{i=1}^N  \nabla_i^2 
        + \sum_{i<j}^{N} V^{\left(\alpha(i),\alpha(j)\right)}(r_{ij})
        + \sum_{i=1}^{N} V_\mathrm{ext}(z_i).
    \end{equation}
The mixture consists of  $N = N_1 + N_2$ atoms, with  $N_1$ ($N_2$) bosons of type 1 (2), $\alpha(i)=1$ for $i\leq N_1$, and $2$ for $i>N_1$. Interatomic potentials $V^{\left(\alpha(i),\alpha(j)\right)}$ are chosen to reproduce the experimental low-energy scattering parameters~\cite{Cabrera}, namely the s-wave scattering length $a$ and effective range $r_{\rm eff}$. Most of the calculations are performed with two potential models, described in the Appendix \ref{app:a}, which reproduce the same values of $a$ and $r_{\rm eff}$ (also given in Appendix \ref{app:a}),  while differing substantially in their shape.   Results obtained with these two models allow us to assess the extent to which the many-body properties are universal in terms of these low-energy parameters and to identify the onset of sensitivity to additional  details of the interaction. The squeezing potential is given as $V_\mathrm{ext}(z_i) = \frac{1}{2}m\omega_z^2z_i^2$, with the frequency $\omega_z = \frac{\hbar}{ma_z^2}$ expressed in terms of the oscillator length $a_z$.

    We rely on the DMC method, which allows one to 
determine the ground-state properties of bosonic systems exactly within statistical noise. In the context of ultracold atoms, it enabled obtaining the 
equation of state of ultradilute bosonic mixtures~\cite{cikojevic2019PRA}. In 
this work, we have proceeded with the same proven second-order DMC 
method~\cite{Boronat:94a}  using a suitable guiding wave function to reduce the 
variance. 
We construct the guiding wave function as a product of 
two-body correlations and  single-particle functions $\phi(z)$ corresponding to the
ground-state of a single atom in the external harmonic potential,
    \begin{align}
    	\Psi(\mathbf{R}) &= 
    	\prod_{i<j}^{N_1} f^{(1,1)}(r_{ij}) 
    	\prod_{i<j}^{N_2} f^{(2,2)}(r_{ij}) 
    	\prod_{i,j}^{N_1,N_2} f^{(1,2)}(r_{ij})\nonumber\\
    	&\times \prod_{i}^{N} \phi(z_{i}),
    	\label{eq:wf}
    \end{align}
where the two-body functions capture the behavior at both large and small distances~\cite{cikojevic2020NJP}
\begin{equation}
	f^{(\alpha,\beta)}(r) =
	\begin{cases}
		f_{2b}(r), & r \leq R_0, \\
		B \exp\left(-\frac{C}{r} + \frac{D}{r^2}\right), & R_0 \leq9 r \leq L/2, \\
		1, & r \geq L/2.
	\end{cases}
\end{equation}
The function $f_{2b}$ is the solution to the two-body problem for the chosen interaction model and is linked to the long-range phononic wavefunction~\cite{Reatto} with coefficients $B$, $C$, and $D$, which are adjusted to ensure the continuity of the wavefunction, its first derivative, and the condition that the derivative is zero at $r = L/2$. The parameter $R_0$ is a variational parameter, and $L = (N/\rho)^{1/2}$ is the size of the simulation box in the direction perpendicular to compression, where $\rho$ is the surface density. 
Simulations are performed in boxes with periodic boundary conditions in the $x$ and $y$ directions. We used $N=400$ for bulk calculations, which was sufficient to keep the finite-size effects under the statistical errors.

For the study of droplets, density functional calculations (DFT) are used, 
relying on the functionals obtained by the DMC calculations. 
The constructed 3D QMC functional was used to study droplets confined even 
further 
than the experimental regime in Ref.~\cite{Cabrera}, from 0.639  to 0.16 $\mu$m 
showing the decrease of the critical number of particles required to form a 
droplet~\cite{Ares}, and persistent difference in prediction with respect to the
3D LHY functional. However, further increasing the confinement, the validity of 
3D functionals is not expected, while previous 
research~\cite{Zin_2022_PRA,Pelayo} suggests that 2D LHY functionals will 
likely not be valid in the crossover regime. 

Within 2D DFT, the 
many-body wave function is built as a product of single particle orbitals 
$\psi_{1,2}(\bf{r}_i)$, which are obtained by solving a Schr\"odinger-like 
equation
\begin{equation}
	\label{eq:dft-2d}
	i\hbar\dfrac{\partial \psi_i}{\partial t} = \left(-\dfrac{\hbar^2}{2m} 
	\nabla^2  + \dfrac{\partial \mathcal{E}_{\rm 
			int}}{\partial \rho_i}\right) \psi_i \ ,
			\end{equation}
where $\mathcal{E}_{\rm int}$ describes the particle correlations.
We solve Eq.~\eqref{eq:dft-2d} using  imaginary-time evolution, with a 
two-dimensional numerical solver based 
	on the Trotter decomposition of the time-evolution 
operator~\cite{chin2009any}. The particles are placed in a box that is adjusted 
according to the droplet size, such that the size of the box and the 
discretization of space do not bias the results. For 3D functionals, we follow 
an analogous procedure, as described in Ref.~\cite{Ares}.

\section{\label{sec:III}Results}
\begin{figure}
	\centering
	\includegraphics[width=0.8\linewidth]{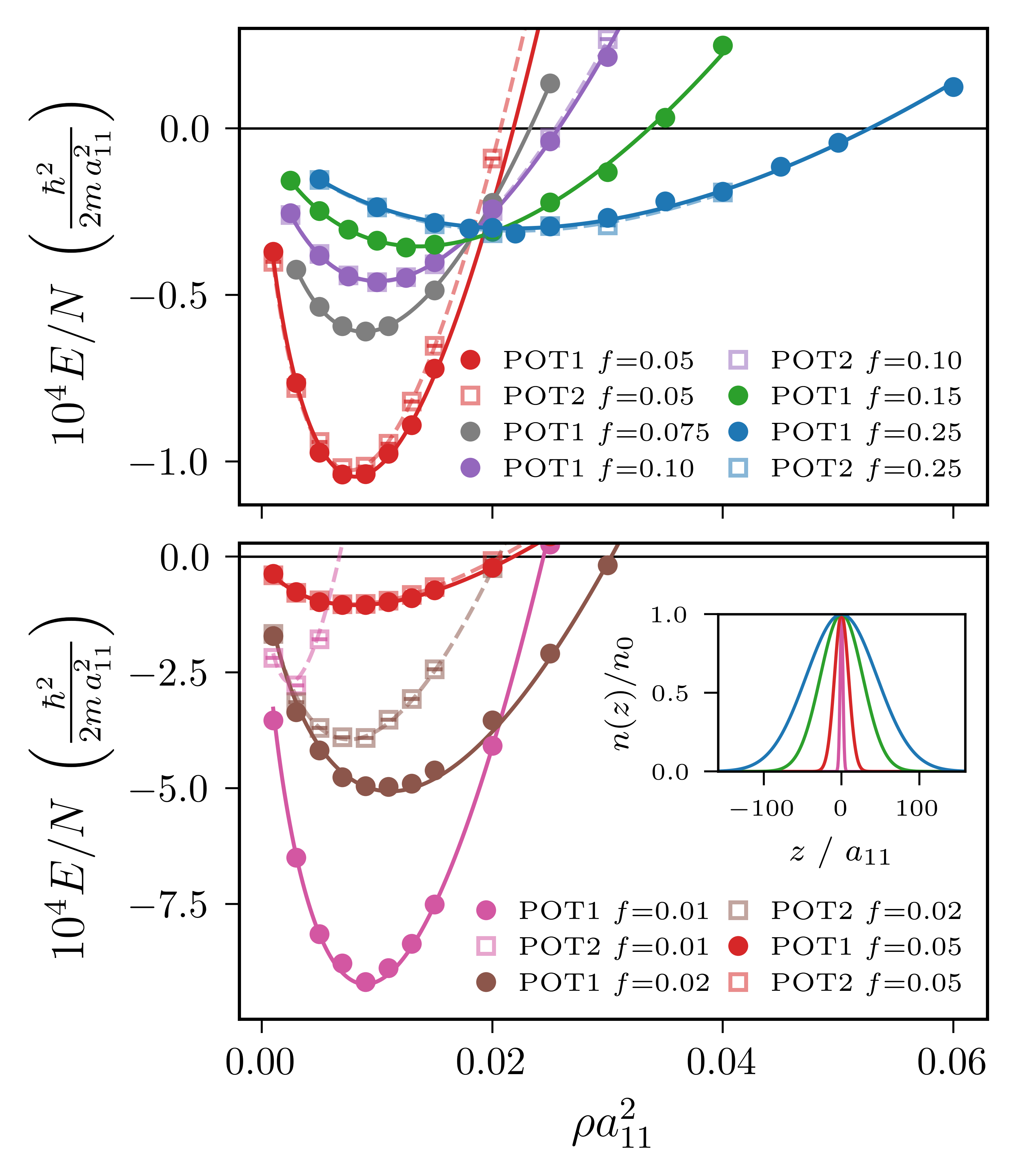}
	\caption{Equation of state of the bulk phase for a magnetic field 
$B=56.337$~G and several harmonic confinements described by the squeezing factor 
$f=a_{z}/0.639~\mu$m. The calculations are performed for two different 
potential models with same scattering length and effective range, showing 
universality up to $f=0.05$. The liquid distribution in the direction of 
squeezing follows has the single-particle gaussian shape, which is shown in the 
inset. From wider, to narrower distributions correspond to $f$=0.25, 0.15, 0.05 
and 0.01.}
	\label{fig:eos}
\end{figure}
In Fig.\ref{fig:eos}, we report the results for the energy per particle as a 
function of the two-dimensional density $\rho$ for a range of squeezing  
described by the harmonic oscillator length of the form $a_{z}=f\times$0.639 
$\mu m$ and the magnetic field $B=56.337$~G, with 
$N_1/N_2=\sqrt{a_{22}/a_{11}}$. The $a_z$ of 0.639 $\mu m$ was used in 
Ref.~\cite{Cabrera} and the range of $f$ from 0.25 to 0.01 allows 
exploration towards the 2D limit, where $a_{z}$ becomes comparable to the s-wave 
scattering lengths. The regimes are indicated in the inset, using the unit of 
length $a_{11}$ to highlight the approach to 2D. With the increase of 
squeezing, the equilibrium energy per particle increases and the equilibrium 
density $\rho_{\rm eq}$ decreases in the range of $f$ from 0.25 to 0.05. In this 
range, the calculations with two different potentials having the same $a$ and 
$r_{\rm eff}$ agree, showing that the description in terms of these two 
parameters is consistent, in line with the previously observed universality of 
ultradilute liquid in 3D~\cite{cikojevic2019PRA,cikojevic2020NJP}. 
However, increasing the confinement even more, we obtain very different results 
for two 
potential models. In the case of POT1 (see Appendix \ref{app:a}), which does not have 
hard core interaction 
at small distances for all $V^{(\alpha,\beta)}(r)$, $\rho_{\rm eq}$  for $f=0.01$ 
is larger by almost an order of magnitude with respect to $f=0.05$, while 
the difference is much smaller in the case of POT2 models, which 
are all strongly 
repulsive at small particle separations. In the latter case, we even observe 
that squeezing from $f=0.02$ to $f=0.01$ causes an increase of equilibrium 
energy per particle. This clearly shows that below the confinement corresponding 
to $f=0.05$ two scattering parameters are not sufficient to describe the 
interaction. The same type of behavior is observed for 
another value of the magnetic field, 
 $B=56.574$~G ($\delta a=a_{12}+\sqrt{a_{11}a_{22}}=-3.2a_0$). 
\begin{figure}
	\centering
	\includegraphics[width=0.9\linewidth]{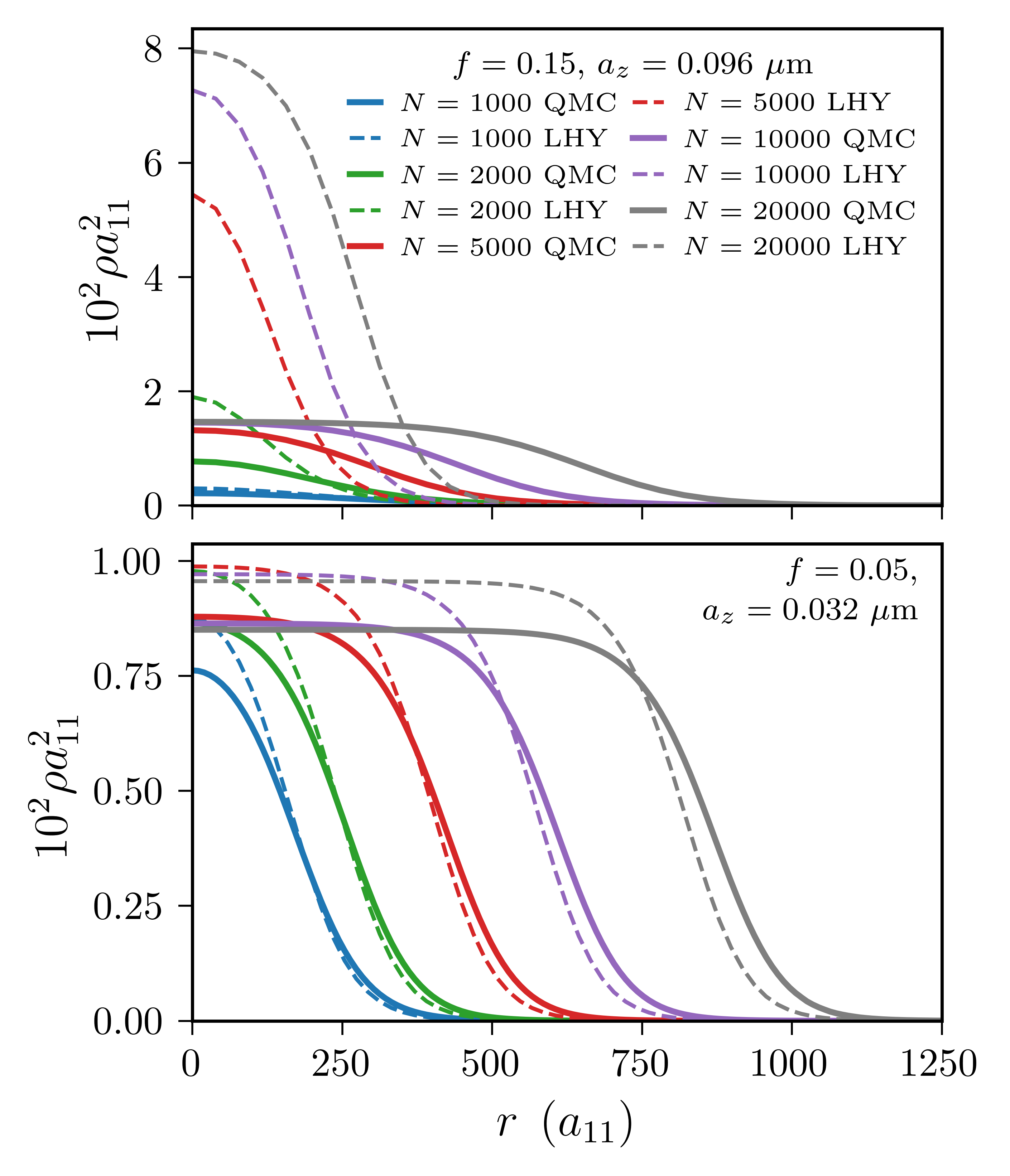}
	\caption{Density profiles of droplets obtained with the 2D QMC and 2D LHY 
functionals compared for two different squeezings $f$=0.15 (top) and 
$f=0.05$ (bottom), for the magnetic field $B=56.337$~G. }
	\label{fig:dens_prof}
\end{figure}
In order to understand this effect, we note that increasing the squeezing leads 
to a confinement-induced increase of the interaction strength, which causes the 
particles to probe the short-range part of the interaction potentials. Since 
that part of the potential differs between the two models, the resulting 
equations of state also differ. 
This effect was previously shown for $^4$He dimers under the same harmonic 
confinement, where a significant enhancement of binding and a decrease in size were predicted when the width of the confining harmonic potential was 
approximately equal to the range of the interparticle 
interaction~\cite{HeliumDimer}.

In the range within universality in the equations of state, $E/N$ 
are fitted to an appropriate functional form, which we use to construct the 2D 
density functional. 
Most studies of bulk helium in 2D used  polynomial expansions around 
the 
equilibrium energy for this purpose. However, we have found that the 
functional form $E/N=\alpha \rho + \beta \rho^\gamma$ performs 
better~\cite{Mur-Petit}. The parameters 
$\alpha$, $\beta$, and $\gamma$ are  obtained by fitting this function to the 
DMC data (see their optimal values in Appendix~\ref{app:b}).   The DMC equation 
of state  is then  included in the interaction part of the 2D 
density functional $\mathcal{E}_{\rm int}= \rho E/N $ in Eq.~\eqref{eq:dft-2d}. The resulting 2D QMC functional is used to study 
droplets, similarly to the study of more weakly confined droplets via 3D QMC and 
3D LHY density functionals~\cite{Ares}. 

In Fig. \ref{fig:dens_prof}, droplet density profiles are shown for two 
different squeezings at $B=56.337$~G, using 2D functionals. Very significant 
difference in density profiles obtained using QMC and LHY functionals is 
observed for $f=0.15$. While QMC functionals predict saturation of central 
density around 10000 atoms, using the LHY functional a droplet with  20000 
atoms has more than 4 times higher central density, but is still not fully 
saturated. The differences are even higher for $f=0.25$. When confinement 
is increased, results approach each other, as 
shown in the lower panel of Fig. \ref{fig:dens_prof}. In that case, the central 
densities of the shown droplets differ by less than 13\%.
\begin{figure}
	\centering
	\includegraphics[width=0.8\linewidth]{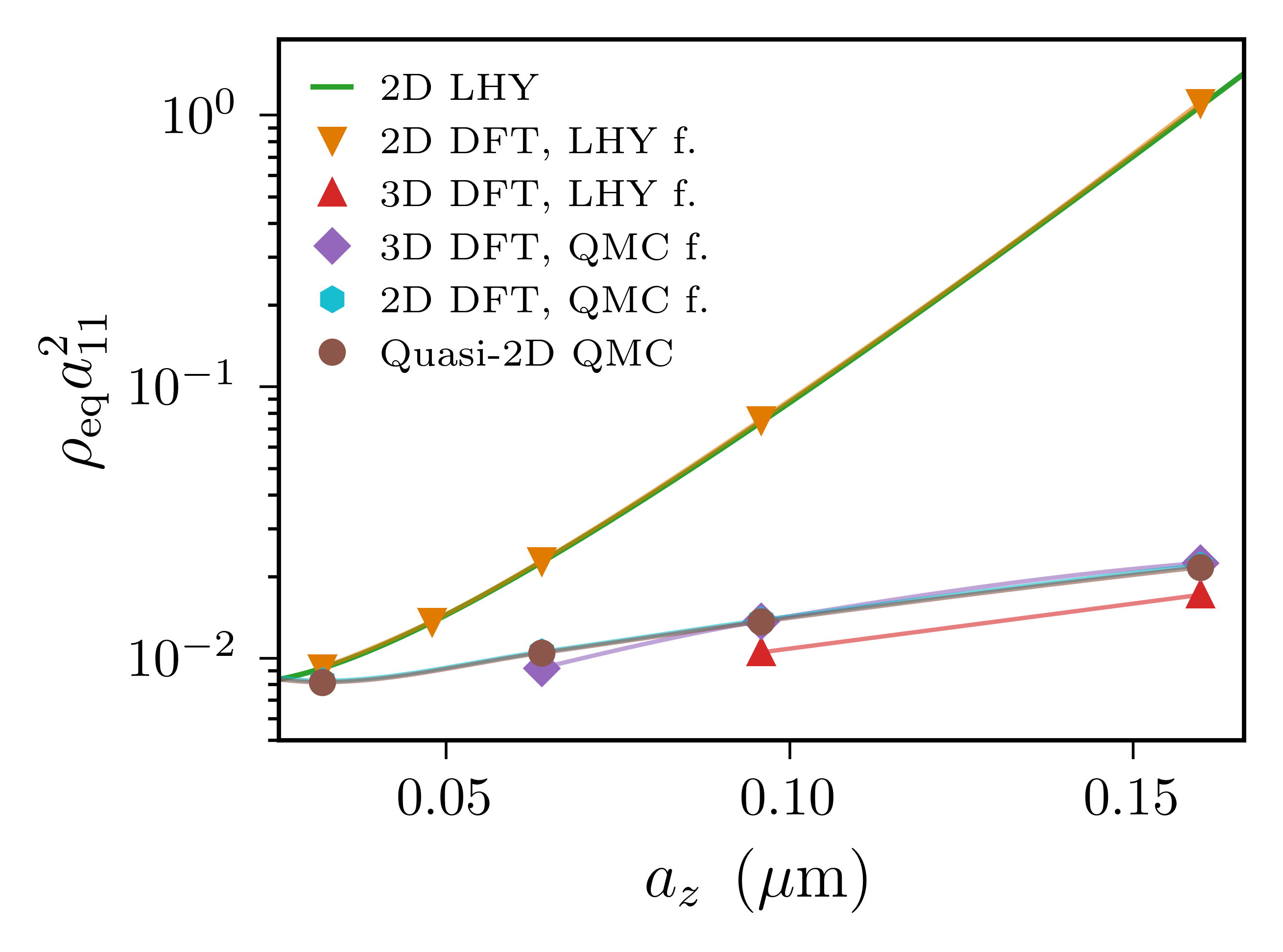}
	\caption{The equilibrium densities of the bulk liquid, obtained with 
quasi-2D QMC calculations are compared to 2D LHY predictions following Ref.~\cite{AP} and to 
saturation densities of droplets studied with DFT functionals, as a function of 
the harmonic oscillator length describing the quasi-2D confinement. We note that the Quasi-2D QMC results overlap the 2D DFT data with the QMC fuctional.}
	\label{fig:eqdens}
\end{figure} 
\begin{figure*}
	\centering
	\includegraphics[width=0.95\linewidth]{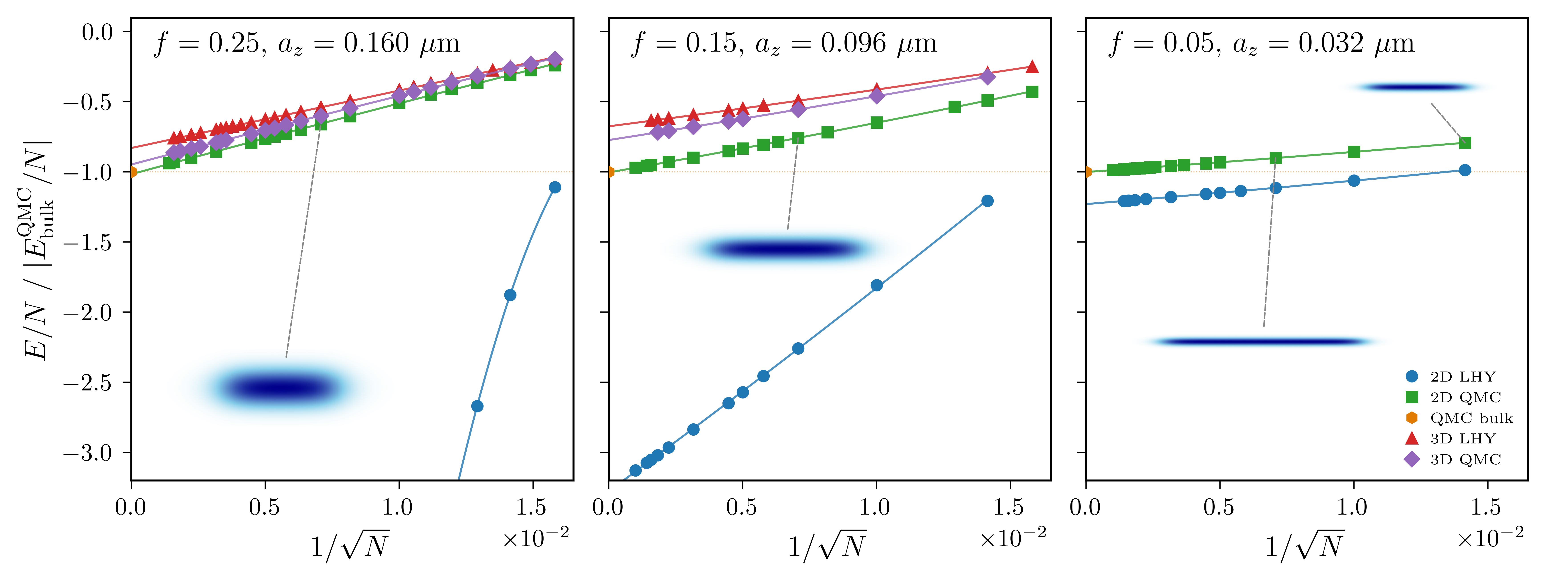}
	\caption{Energy per particle of the droplets as a function of $N^{-1/2}$ calculated with 2D and 3D QMC and LHY functionals, for three different squeezings. The dots at zero $x$ axis correspond to the equilibrium energy in quasi-2D bulk DMC calculation. The insets show the cut through the 3D droplet profile at $x=0$ for droplets of size $N=20000$ and $N=5000$ (for the strongest squeezing only). The $z$ coordinate has been multiplied by a factor of 10 for clarity. }
	\label{fig:liquid_drop}
\end{figure*}
  
  In Fig. \ref{fig:eqdens}, we compare the equilibrium densities obtained from 
QMC quasi-2D bulk calculations with DFT results using both 2D and 3D 
functionals, and with the 2D LHY prediction from Ref.~\cite{AP}, which is obtained using optimal $N_1/N_2$ for 2D, as explained in 
the Appendix \ref{app:c}. The difference between the latter and the 2D LHY calculation using the 3D  ratio $N_1/N_2=\sqrt{a_{22}/a_{11}}$ is negligible.
From Fig. \ref{fig:eqdens} it is clear that with increasing $a_z$ the 
quasi-2D DMC results increasingly depart from the LHY predictions, as already 
indicated by the profiles shown in Fig.~\ref{fig:dens_prof}. The saturation 
densities of droplets obtained with the 2D QMC functional in the limit  $N \to 
\infty$  reach the quasi-2D bulk DMC value, demonstrating 
functional's reliability  for large droplets. Surprisingly, the predictions of 
the saturation density using 3D QMC functional almost coincide with those using 
2D QMC functional down to $a_z=0.1~\mu$m, while those obtained using 3D LHY 
differ by up to 23\%. In addition, as $a_z$ is decreased further, the 3D functionals no longer predict self-bound droplets.
The results mentioned previously are qualitatively the same for the other magnetic field studied, $B$ = 56.754 G, at which $|\delta a|$ has a smaller value. There, the 2D LHY results differ from the 2D QMC by 22\% at $a_z=0.048~\mu$m ($f=0.075$), while the same difference at the field $B$ = 56.337 G appears at $a_z=0.037~\mu$m ($f=0.058$). 
  
  Recently, path integral Monte Carlo (PIMC) calculations have been used to 
study the formation of symmetric droplets  ($a_{11}=a_{22}=a$) in strong 
harmonic confinement~\cite{Spada}. Although we cannot compare our results 
directly, we can use the 3D QMC functional for the symmetric 
mixture~\cite{cikojevic2019PRA}, as well as 2D and 3D LHY functionals. For  
$a_{12}=-1.1a$, $a_z\simeq 31.6a$, taking $a=a_{11}$ for $^{39}$K at $B=56.337$~G, 
one gets  $a_z\simeq 0.13\, \mu$m, which is a regime in which we still find 
agreement with 3D functionals. The reported central droplet density in 
Ref.~\cite{Spada} $na^2\simeq 0.0025$ is indeed found to be six times lower than 
the predictions using 2D LHY functionals $na^2\simeq 0.015$. On the other hand, 
as expected, we get much closer results with 3D functionals,  $na^2\simeq 
0.0033$ with 3D QMC  and $na^2\simeq 0.0035$ with 3D LHY functional, assuming a 
droplet size of 16000 atoms. The difference between 3D QMC functional results (obtained with interactions reproducing the experimental values of the $s$-wave scattering length and effective range) and PIMC calculations (obtained with contact interactions) represents the likely size of finite range contributions. It is interesting to note that in
Ref.~\cite{Spada} 
excellent agreement was obtained between PIMC calculations in quasi-2D and in 
strictly 2D, for the correct logarithmic form of the 2D coupling strength,  
however, in both cases without considering finite range effects. This 
highlights the importance of the QMC approaches which can take into account full 
correlations, beyond those accessible by the LHY term.

Besides equilibrium densities, it is interesting to compare the predictions 
for the droplets' energies per particle obtained using different
functionals. In 2D systems, previous research on helium 
droplets~\cite{Krishnamachari,Sarsa,graphene2} has shown that the energy per 
particle is well 
described by a liquid drop model, according to which $E/N=E_b +E_lx +E_cx^2$, 
with $x=N^{-1/2}$, and $E_b,\, E_l,\, E_c$ the fitting parameters. $E_b$ corresponds 
to the bulk energy per particle, $E_l$ is the line (surface) energy and $E_c$ is 
the so-called curvature energy. The same form of scaling with $N$ was shown in the variational study of 2D Bose-Bose droplets~\cite{examilioti_ground_2020}. 
In Fig. \ref{fig:liquid_drop} the energies per particle are presented in units 
of the absolute values of the equilibrium energy obtained by the bulk quasi-2D 
DMC calculations. In order to compare energies with 2D and 3D functionals, 
energy contribution of the single particle in the harmonic oscillator potential 
is subtracted from the total energy in 3D, leaving the self-binding energy 
contribution. The energies per particle calculated using 2D QMC functional 
clearly extrapolate to the bulk DMC value for all studied squeezings. For the 
largest considered $a_z$, 3D and 2D QMC functionals give similar predictions, 
while 3D LHY functional predicts up to 20\% less self-binding than 2D QMC 
results. 
The most striking feature is the huge difference in the predicted energies per 
particle with the 2D LHY functional, with more than 10 times stronger 
self-binding  for $N>10^4$ with respect to 2D QMC values. These differences are 
reduced when approaching the 2D limit, reaching, in the case of smaller 
droplets, up to 25\%  with respect to 2D QMC values for $f=0.05$ ($a_z$ = 
9.07$a_{11}$). For this confinement, no predictions using 3D functionals are 
given, because they result in positive $E/N$. It is important to notice that across
the entire squeezing range the droplets are very flat, as illustrated by several 
droplet profiles in the $x-z$ plane,  in which the $z$ coordinates have been 
multiplied by a factor of 10 for better readability. 

In order to differentiate the finite-range and beyond LHY correlations from the dimensionality effects, we would need to compare 2D QMC functional results with the
functional that includes beyond MF effects in the crossover from 3D to 2D dimensions~\cite{Ilg,Zin18,Zin_2022, Zin_2022_PRA, Pelayo}. Such functionals are however unavailable in harmonic confinement for realistic asymmetric $^{39}$K mixtures. The closest case is a recent study~\cite{Pelayo} of quasi-2D droplets, in a tightly confined box
of size $L_z$.  For $\delta a=-5.5a_0$, it was found~\cite{Pelayo} that when $L_z=0.491~\mu$m the
energies using quasi-2D LHY correspond to those using 3D LHY term, while already 
at $L_z=0.087~\mu$m the 2D LHY functional is not accurate enough and needs to be
replaced by the quasi-2D LHY.   The range of confinement for which either the 2D or
3D approach fail is in good agreement with our results if we match the 
ground-state root-mean-square size of droplets in the direction of confinement, 
arriving at $L_z\approx 2.8 a_{z}$.  This suggests that the dimensionality effects are more significant than beyond LHY correlations and finite-range effects, which are not taken into account in the approach of Ref.~\cite{Pelayo}. 

\begin{figure}
	\centering
	\includegraphics[width=0.95\linewidth]{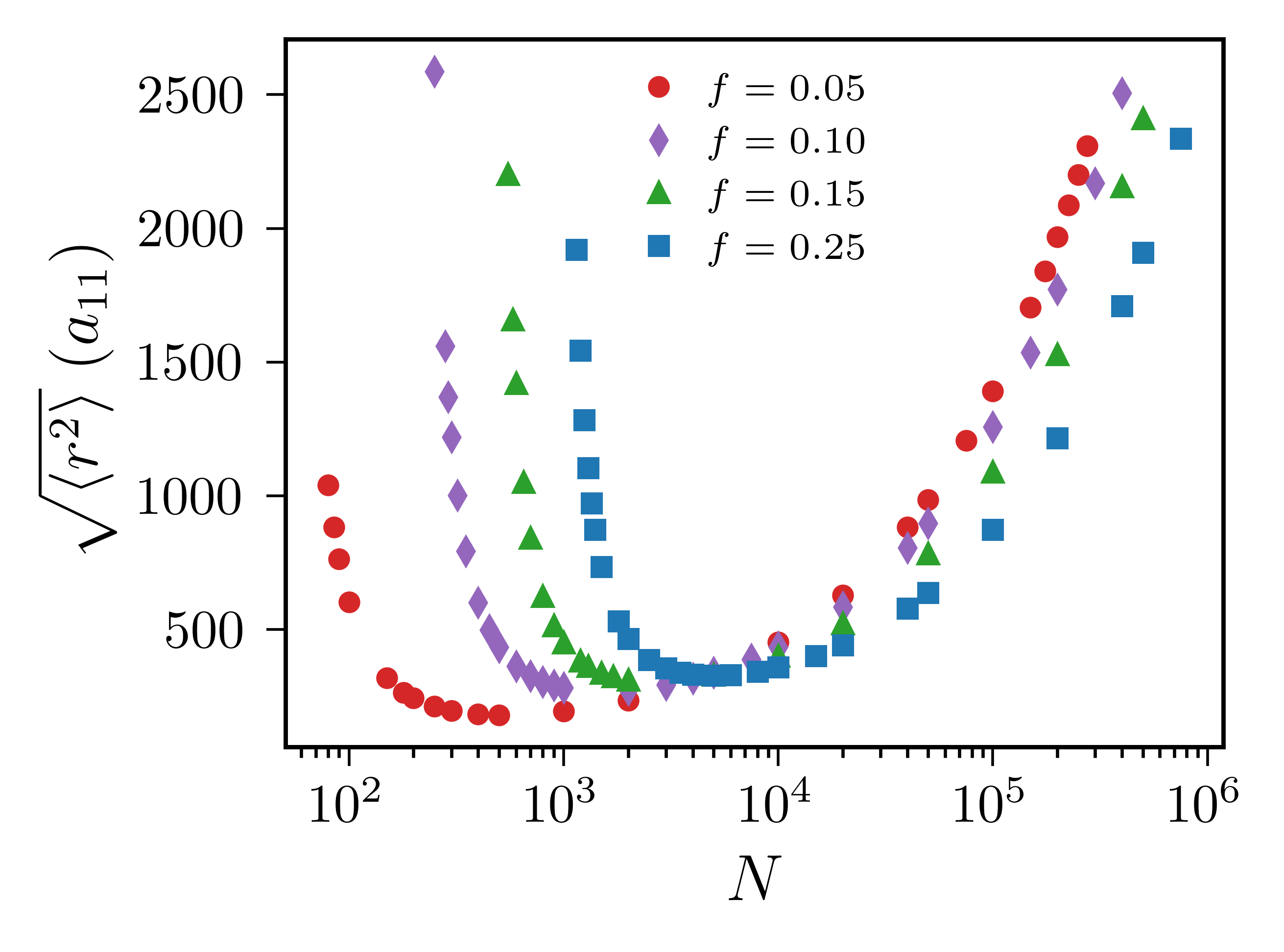}
	\caption{Root mean square radius of droplets for different confinements at $B=56.337$ G.}
	\label{fig:rms}
\end{figure}

Using the 2D QMC functional and following the procedure from Ref.~\cite{Ares} we were able to determine the critical number of atoms $N_c$ required to form a self-bound droplet as the droplets are squeezed towards 2D. The results, shown in  Table \ref{tab:CritNumber}, demonstrate a continued reduction of $N_c$ with increasing confinement, as previously observed in Ref.~\cite{Ares} with 3D functionals in weaker confinements. The predictions obtained using 2D and 3D QMC functionals for $a_z=0.160$ $\mu$m agree quite well, with the 2D functionals predicting slightly lower $N_c$ caused by slightly stronger self-binding. For the strongest squeezing where universality of the equation of state was observed, DFT predicts $N_c$ below 100 atoms and it would be worthwhile to study those droplets with direct quantum Monte Carlo calculations in order to determine how well DFT performs at such low particle numbers. It is interesting to note that, despite the very narrow confinements, we still do not reach the mathematical 2D limit, where binding is expected for all $N \ge 2$~\cite{AP}.
\begin{table}[h]
    \caption{\label{tab:CritNumber} Critical number of atoms $N_c$ needed to form a bound droplet for two different magnetic fields and various confinement strengths. The second line for $f=0.25$ are results obtained in~\cite{Ares}.}
        \begin{ruledtabular}
            \begin{tabular}{cccc}
                $f$ & $a_z$ ($\mu$m) & $B = 56.337$ G & $B = 56.574$ G \\\hline
             \multirow{2}{*}{0.25} & \multirow{2}{*}{0.160} 
                & 1150(50) & 1900(50) \\
                & & 1200(30) & 2250(50) \\\hline
                0.15 & 0.096 & 570(30) & 800(30)\\ 
                0.10 & 0.064 & 280(20) & 440(30)\\ 
                0.05 & 0.032 & 84(10) & 100(10)\\ 
            \end{tabular}
        \end{ruledtabular}
\end{table}
In addition to $N_c$, it is possible to experimentally measure the size of the droplets, which we show in Fig.~\ref{fig:rms} for different confinements at $B=56.337$ G.
The root mean square radius shows a non-monotonic dependence on the number of particles in the droplet. Close to $N_c$ the radius is very large, it reduces as $N$ increases and reaches the minimum before increasing again.  
When droplets are close to $N_c$, increasing the squeezing for a fixed $N$ strongly reduces the spatial extent of the droplet. The opposite is observed for droplets whose central density is saturated. That is, the size of the saturated droplets grows proportionally to $\sqrt{N}$, which was also observed in 2D helium droplets~\cite{Mur-Petit}. The proportionality constant is in good agreement with $1/\sqrt{2\pi \rho_0}$, where $\rho_0$ is the density of the saturated droplet. As discussed before, more strongly confined droplets have lower central density.

\begin{figure}
	\centering
	\includegraphics[width=0.95\linewidth]{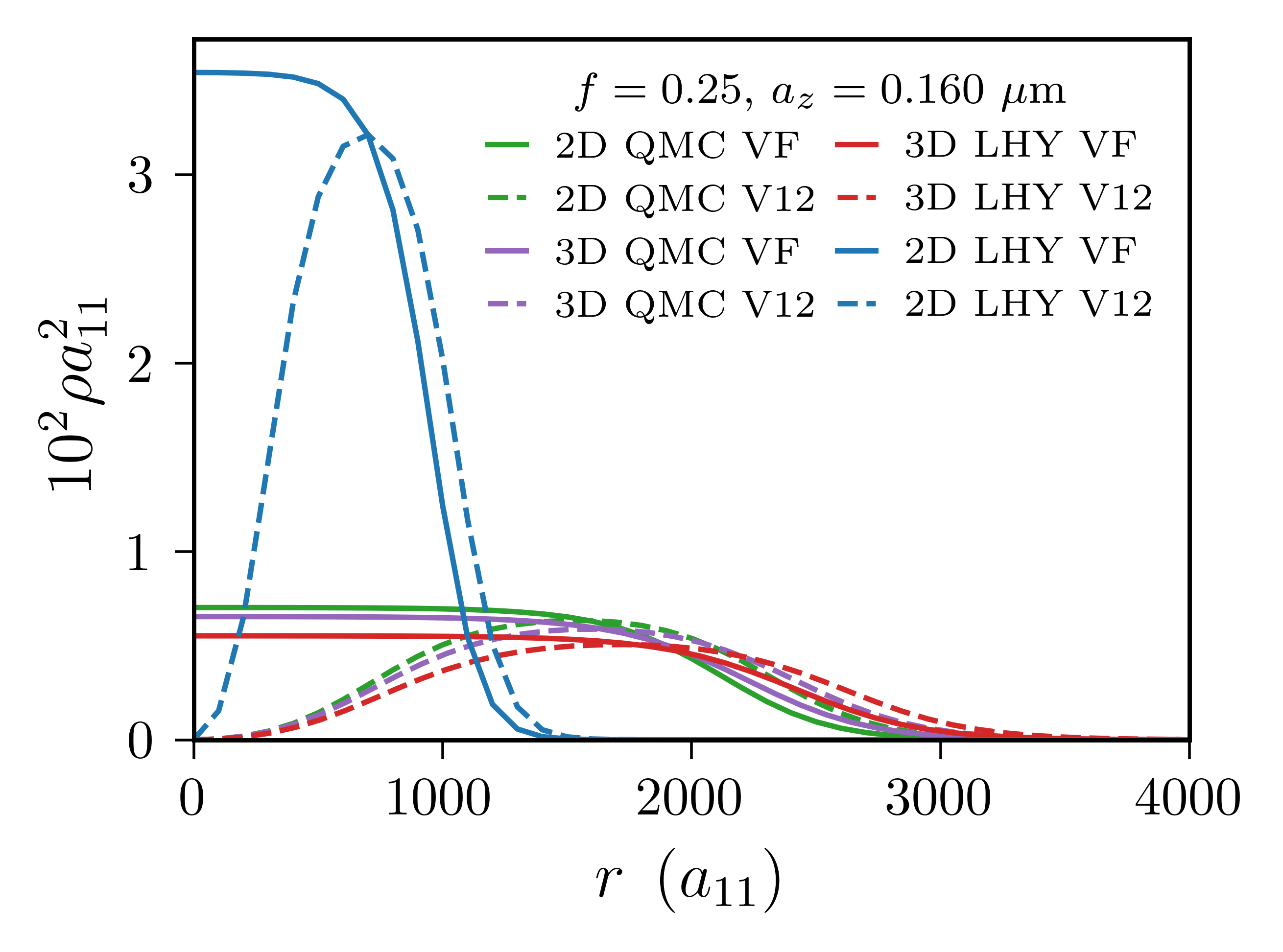}
	\caption{Density profiles of droplets with 100000 atoms in a magnetic field 
$B=56.574$~G  with $f=0.25$ and angular velocity $\Omega=2\pi\times 
10$ Hz, without vortex (VF) and with vortex in both components (V12), calculated 
using different density functionals. Droplet with vortex calculated using 2D LHY functional is metastable. }
	\label{fig:vortex}
\end{figure}

The significant differences in predictions of different functionals are not constrained only to density
profiles and energies, but will appear in the study of excitations as well. It 
is very common to study vortices using the 2D LHY 
functional~\cite{Li2018,tengstrand_rotating_2019,Kartashov19,kartashov_2020,
examilioti_ground_2020,Dong21,Sturmer21,nikolaou_rotating_2023-1,gu_self-bound_2023,
nikolaou_rotating_2024,Wang2025}  which are then generalized to a quasi-2D 
situation using the mapping of the 2D scattering length to the particular 
harmonic confinement~\cite{Petrov2001}. Using a 3D LHY functional we have 
studied the vortex formation in Bose-Bose mixtures~\cite{poparic} and predicted 
the regions of stability of empty and filled vortices. In order to demonstrate 
the sensitivity of the obtained results on the employed  approach, we show in 
Fig. \ref{fig:vortex} the droplet profiles obtained using four different functionals, 2D and 3D versions of both QMC and LHY 
functionals, for squeezing factor $f=0.25$ ($a_z=0.16\,\mu$m) in magnetic field 
$B=56.574$~G and angular velocity $\Omega=2\pi\times 10$ Hz.  In order to compare the profiles from 2D and 3D calculations, those obtained in the 3D calculations are integrated in the $z$ direction.
For this magnetic field ($\delta a=-3.2$a$_0$), 2D LHY in vortex-free droplets 
predicts more than 4 times higher central densities than other functionals, and 
it finds that the vortex-free state is energetically more favorable than the 
metastable vortex state. In addition, it produces significantly smaller vortex 
cores. On the other hand, the 2D QMC functional predicts a slightly smaller 
vortex core than the 3D QMC functional, which relies on the local density 
approximation in the direction of squeezing. Importantly, both QMC functionals 
find the vortex state as more favorable, supporting  previous 
findings~\cite{poparic}. Furthermore, calculating the critical number of atoms for vortex hosting at this angular velocity, we find that the QMC functionals predict somewhat larger numbers, $N_\text{c,v}^\text{3D QMC} \approx 60 000$ and $N_\text{c,v}^\text{2D QMC} \approx 70 000$, as opposed to the $N_\text{c,v}^\text{3D LHY} \approx 53 000$ found in our earlier work. The 2D LHY functional predicts an almost ten-fold larger number of $N_\text{c,v}^\text{2D LHY} \approx 480 000$. This remains consistent with our previous finding that the critical number is larger in more self-bound droplets, as the 2D LHY functional has several times stronger binding, while among the other three, 3D LHY has the weakest and 2D QMC has the strongest self-binding. 

\section{\label{sec:IV}Conclusions}

In conclusion, we have shown that quantum Monte Carlo is able to provide 2D 
functionals that can accurately predict droplet energies and density profiles in 
the crossover from 3D to 2D.  Besides capturing the dimensionality effects, these functionals go beyond LHY corrections and are 
universal with respect to two parameters, the scattering length and effective 
range, up to the harmonic oscillator length of roughly 10 $a_{11}$, when they 
also approach the 2D LHY prediction. For even narrower confinements, 
details of the interaction potential beyond $r_{\rm eff}$ and $a$ become 
important. This signifies a very narrow range of applicability of 2D LHY 
functionals to confined droplets, which is important to realize when guiding the experiments.  The 
finding has important implications for the study of vortices that have still not 
been detected in these droplets, as well as excitations. Obtained 2D QMC 
functionals are a simple alternative, which is computationally as efficient as 
effective single-component LHY 2D functionals. A similar approach could become 
very relevant for strongly interacting self-bound droplets of dipolar molecules 
where contributions beyond LHY are very important and where it is also 
challenging to describe accurately the crossover from 3D to 2D. 

We have used the functionals to predict experimentally relevant information, the critical atom numbers for self-bound states, $N_c$, and droplet sizes in the confinement range studied, demonstrating the expected reduction of $N_c$ with squeezing and the nonmonotonic dependence of the droplet radius on the number of particles.

The quantitative confinement scales reported in the present work are specific to the $^{39}$K mixture and interaction parameters considered here.
At the same time, the physical mechanism underlying the crossover is not specific to the
$^{39}$K mixture. We expect that it results from the competition of the transverse confinement length, interparticle interaction, and density regimes. While determining the details of this dependence is beyond the scope of our work, we expect that mixtures having comparable dimensionless ratios of these quantities will show  analogous behavior, starting with 3D to quasi-2D behavior, and then evolving to a regime where a 2D description becomes appropriate. Also, that under sufficiently strong confinement increased sensitivity to details of the interaction potential, beyond the $s$-wave scattering length and effective range, will emerge.

\begin{acknowledgments}
We thank F. Mazzanti and R. Zillich for discussions.
We acknowledge financial support from the Croatian Science Foundation projects IP-2022-10-6144  and DOK-NPOO-2023-10-5811, European Union (NextGenerationEU) under the Croatian Recovery and Resilience Plan 2021–2026 (NRRP), through the University of Split institutional project IP-UNIST-46 and from Ministerio de Ciencia e
Innovaci\'on MCIN/AEI/10.13039/501100011033
(Spain) under Grant No. PID2023-147469NB-C21.
	The computational resources of UniST-Phy server at the University of Split and supercomputers Supek and Padobran at SRCE in Zagreb were used.
\end{acknowledgments}

\appendix
\section{\label{app:a}Interaction potential models}
In our calculations we considered two sets of interaction potential models, POT1 and POT2, with the same scattering lengths $a$ and effective ranges $r_{\rm eff}$. The first model POT1 is given by 
    \begin{align}
		V^{(1,1)}(r) &= 
		\begin{cases} 
			-V_0, & r \leq R_0, \\
			V_1, & R_0 < r \leq R_1, \\
			0, & R_1 < r,
		\end{cases} \label{eq:V11}\\
		V^{(1,2)}(r) &= 
		\begin{cases} 
			-V_0, & r \leq R_0, \\
			0, & r > R_0,
		\end{cases} \label{eq:V12}\\
		V^{(2,2)}(r) &= V_0 \left[ \left( \frac{r_0}{r} \right)^{10} - \left( \frac{r_0}{r} \right)^6 \right]. \label{eq:V22}
	\end{align}

    The second model POT2 has the same form for $V^{(2,2)}$, but different form for $V^{(1,1)}$ and $V^{(1,2)}$, 
    \begin{align}
		V^{(1,1)}(r) &= V_0 \left[ \left( \frac{r_0}{r} \right)^{10} - \left( \frac{r_0}{r} \right)^6 \right] \nonumber\\
            &+V_1 \exp\left({-\frac{(r-r_1)^2}{2r_2^2}}\right),
        \label{eq:Vb11}\\
		V^{(1,2)}(r) &= V_0 \left[ \left( \frac{r_0}{r} \right)^{10} - \left( \frac{r_0}{r} \right)^6 \right]. \label{eq:Vb12}
	\end{align}

    Parameters for the potentials POT1 and POT2 are given in the Tabs.~\ref{tab:V11}, \ref{tab:V12}, \ref{tab:V22}, \ref{tab:Vb12} and \ref{tab:Vb11}, where the lengths are given in units of scattering lengths of $V^{(1,1)}$, which are $a_{11}=66.619\ a_0$ and $a_{11}=74.118\ a_0$, respectively for $B=56.337$~G and $B=56.574$~G, while energy unit is $E_1=\hbar^2/(2 m a_{11}^2)$, i.e. $1\ {\rm m}E_1 = 10^{-3}E_1$.

    \begin{table}[h!]
    \caption{\label{tab:V11} Parameters for $V^{(1,1)}$ in POT1 given by Eq.~\eqref{eq:V11}.}
        \begin{ruledtabular}
            \begin{tabular}{cccccc}
                $B ({\rm G})$ & $R_0 (a_{11})$ & $R_1 (a_{11})$ & $V_0 ({\rm m}E_1)$ & $V_1 ({\rm m}E_1)$ & $r_{\rm eff} (a_{11})$\\\hline
                56.337 & 3.549036 & 6.214063 & 90.61211 & 59.23397 & -17.34145\\
                56.574 & 3.541946 & 6.176794 & 69.64813 & 54.95676 & -15.45753\\
            \end{tabular}
        \end{ruledtabular}
    \end{table}
    \begin{table}[h!]
    \caption{\label{tab:V12} Parameters for $V^{(1,2)}$ in POT1 given by Eq.~\eqref{eq:V12}.}
        \begin{ruledtabular}
            \begin{tabular}{ccccc}
                $B ({\rm G})$ & $r_0 (a_{11})$ & $V_0 ({\rm m}E_1)$ & $a_{12} (a_{11})$ & $r_{\rm eff} (a_{11})$ \\\hline
               56.337 & 4.629644 & 40.10681 & -0.801363 & 15.35055 \\
               56.574 & 4.162949 & 49.50344 & -0.718827 & 13.82660 \\
            \end{tabular}
        \end{ruledtabular}
    \end{table}
    \begin{table}[h!]
    \caption{\label{tab:V22} Parameters for $V^{(2,2)}$ in POT1 and POT2 given by Eq.~\eqref{eq:V22}.}
        \begin{ruledtabular}
            \begin{tabular}{ccccc}
                $B ({\rm G})$ & $r_0 (a_{11})$ & $V_0 (E_1)$ & $a_{22} (a_{11})$ & $r_{\rm eff} (a_{11})$ \\\hline
               56.337 & 1.279465 & 5.315555 & 0.515904 & 8.827617 \\
               56.574 & 1.147111 & 6.719496 & 0.457311 & 8.239469 \\
            \end{tabular}
        \end{ruledtabular}
    \end{table}
    \begin{table}[h!]
    \caption{\label{tab:Vb12} Parameters for $V^{(1,2)}$ in POT2 given by Eq.~\eqref{eq:Vb12}.}
        \begin{ruledtabular}
            \begin{tabular}{ccccc}
                $B ({\rm G})$ & $r_0 (a_{11})$ & $V_0 (E_1)$ & $a_{12} (a_{11})$ & $r_{\rm eff} (a_{11})$ \\\hline
               56.337 & 0.9844272 & 28.87534 & -0.801363 & 15.35055 \\
               56.574 & 0.884763  & 35.72980 & -0.718827 & 13.82660 \\
            \end{tabular}
        \end{ruledtabular}
    \end{table}
    \begin{table}[h!]
    \caption{\label{tab:Vb11} Parameters for $V^{(1,1)}$ in POT2 given by Eq.~\eqref{eq:Vb11}.}
        \begin{ruledtabular}
            \begin{tabular}{cccccc}
                $B ({\rm G})$ & $r_0 (a_{11})$ & $r_1 (a_{11})$ & $r_2 (a_{11})$ & $V_0 (E_1)$ & $V_1 (E_1)$ \\\hline
                56.337 & 0.63408 & 0.73302 & 7.098 & 0.6774 & 0.01760 \\
                56.574 & 0.33190 & 0.38550 & 6.229 & 0.3510 & 0.05553  \\
            \end{tabular}
        \end{ruledtabular}
    \end{table}

\section{\label{app:b}Equations of state}
The obtained parameters for the equation of state for two magnetic fields and several confinements, using POT1 models for interactions, are given in table \ref{tab:parameters}. They are obtained for the $N_1/N_2=\sqrt{a_{22}/a_{11}}$, which we verified was optimal in the for the given range of confinements.
\begin{table}[h]
    \caption{\label{tab:parameters} Parameters $\alpha$, $\beta$ and $\gamma$ of the equation of state as a function of magnetic field $B$ (in G) and confinement (in $\mu$m). $\alpha$ is given in units of $\hbar^2/2m$, $\beta$ in $\hbar^2 a_{11}^{2\gamma-2}/2m$ and $\gamma$ is dimensionless.}
        \begin{ruledtabular}
            \begin{tabular}{ccccc}
                $B$ (G) & $a_z$ ($\mu$m) & $\alpha\!\left(\frac{\hbar^2}{2m}\right)$ & $\beta\!\left(\frac{\hbar^2 a_{11}^{2\gamma-2}}{2m}\right)$ & $\gamma$ \\\hline
             \multirow{6}{*}{56.337}
                & 0.032 & $-0.446218$ & $0.499704$ & $1.029605$ \\
                & 0.038 & $-0.412209$ & $0.452152$ & $1.024276$ \\
                & 0.048 & $-0.104690$ & $0.136468$ & $1.070502$ \\
                & 0.064 & $-0.039203$ & $0.063440$ & $1.131546$ \\
                & 0.096 & $-0.014410$ & $0.030440$ & $1.220469$ \\
                & 0.160 & $-0.006610$ & $0.014409$ & $1.264896$ \\\hline
             \multirow{6}{*}{56.574}
                & 0.032 & $-0.530539$ & $0.592275$ & $1.025460$ \\
                & 0.048 & $-0.067881$ & $0.108941$ & $1.105680$ \\
                & 0.064 & $-0.042470$ & $0.068561$ & $1.106460$ \\
                & 0.096 & $-0.012485$ & $0.029842$ & $1.197400$ \\
                & 0.160 & $-0.005150$ & $0.013720$ & $1.237818$ \\
            \end{tabular}
        \end{ruledtabular}
\end{table}
\section{\label{app:c}2D LHY expressions}
Based on the derivations from Ref. \cite{AP}, the 2D equilibrium density and 
energy density are given by 
\begin{eqnarray}
   \label{eq:rhoeq}
   \rho_0&=&\frac{\Delta e^{-3/2}}{\sqrt{g_{11}}\sqrt{g_{22}}} ,\\
   \mathcal{E}&=&\frac{\hbar^2}{8m\pi}g_{11}g_{22}\rho^2\left(\ln{\frac{\rho}{\rho_0}}-1 \right),
   \label{eq:enLHY}
\end{eqnarray}
where 
\begin{equation*}
\Delta=\sqrt{\epsilon_{12}\sqrt{\epsilon_{11}\epsilon_{22}}}\exp{\left[\frac{-\ln^2(\epsilon_{11}/\epsilon_{22})}{4\ln(\epsilon_{11}\epsilon_{22}/\epsilon_{12}^2)}\right]}~,
\end{equation*}
$g_{ij}=4\pi/\ln(\epsilon_{ij}/\Delta)$ and $\epsilon_{ij}$ are adjusted to a particular confinement by 
$\epsilon_{ij}=(A/\pi a_z^2)\exp{[\sqrt{2\pi}a_z/a^{3D}_{ij}]}$. The expressions~\eqref{eq:rhoeq} and~\eqref{eq:enLHY} assume $N_1/N_2=\sqrt{g_{22}/g_{11}}$.

\end{document}